%% file: DCC_occ-paper.tex
\documentclass[a4paper, authoryear]{article}
\usepackage{techRepsPDF}
\usepackage{graphicx}
\usepackage{color}
\usepackage{array}
\usepackage[lowtilde]{url}
\usepackage{authblk}

\usepackage[linesnumbered,ruled,vlined,]{algorithm2e}

\SetCommentSty{mycommfont}
\newcommand{\eq}{\leftarrow}
\newcommand{\ie}{\emph{i.e.}}
\newcommand{\eg}{\emph{e.g.}}

\def\ptitle{A Practical Algorithm for Packing Tubes and Boxes}
\def\pnumber{DCC-2016-02}
\def\pauthor{Jo\~ao Pedro Pedroso and Jo\~ao Nuno Tavares and Jorge Leite}

\trtitle{\ptitle}
\trnumber{\pnumber}
\trauthor{\pauthor}

\begin{document}

\mkcoverpage

\title{A Practical Algorithm for Packing Tubes and Boxes}

\author[1]{Jo\~ao Pedro Pedroso\thanks{jpp@fc.up.pt}}
\author[2]{Jo\~ao Nuno Tavares\thanks{jntavar@fc.up.pt}}
\author[3]{Jorge Leite\thanks{jorge.leite@fersil.com}}
\affil[1]{INESC TEC and Faculdade de Ci\^encias, Universidade do Porto, Portugal}
\affil[2]{Centro de Matem\'atica and Faculdade de Ci{\^e}ncias, Universidade do Porto, Portugal}
\affil[3]{FERSIL -- Freitas \& Silva, S.A., Cesar, 3700 Oliveira de Azem{\'e}is, Portugal}

\date{September 2016}

\maketitle

\begin{abstract}
In this paper we describe a method for packing tubes and boxes in containers.  Each container is divided into parts (holders) which are allocated to subsets of objects.  The method consists of a recursive procedure which, based on a predefined order for dealing with tubes and boxes, determines the dimensions and position of each holder.   Characteristics of the objects to pack and rules limiting their placement make this problem unique.  The method devised provides timely and practical solutions.
\end{abstract}

\section{Introduction}
\label{sec:intro}

Practical packing problems depend deeply on the shape and other characteristics of the products involved.  In the tube industry these problems are very complex and difficult, not only due to the variety of possible shapes --- tubes of many different radii and lengths, boxes, cylinders --- but also because of the requirement to quickly obtain a solution.

Companies in this industry face several challenges in the optimisation of load for product delivery.  The ``standard'' way of delivering merely fixed quantities ordered by customers is not satisfactory, as it usually leads to unused space inside containers.  Regular customers prefer to completely use the container, so as to reduce the frequency of deliveries.  On the other hand, the extreme complexity of packing tubes of different dimensions, often of several lengths, and mixed with boxes and bags, makes the task of planning a delivery very time consuming.  An additional difficulty is that straightforward methods are clearly suboptimal.  

We are hence in a situation where quantities ordered are often not fixed, as both the company and frequent customers want to fully utilise containers capacity.  For this purpose, interaction between the logistics and commercial departments is essential, as customers usually want to be informed of what can be delivered.  This calls for an interactive, responsive system for supporting operational decisions in container loading and packing.  We propose a heuristic method for packing tubes and boxes together, based on a recursive partition of the container into smaller parts.  

Three-dimensional packing has recently been studied under several different perspectives.  The problem of allocating a given set of three-dimensional rectangular items to the minimum number of identical finite bins without overlapping has been addressed with tabu search in~\cite{Lodi2002410}; here, items are packed in several layers, the floor of the container being the first.  A heuristic method for the situation where there is no requirement for packed boxes to form flat layers, keeping track of empty space seen from different perspectives and using a look-ahead scheme for positioning, is presented in~\cite{Lim2003471}.  Another tabu search for three-dimensional orthogonal bin packing is presented in~\cite{Crainic2009744}; in this case it is a two-level approach, separating the search for the optimal number of bins from optimising of the accommodation of items within bins. 

In our problem, minimising the number of containers and maximising space usage are conflicting objectives, as often arises in packing (see, \eg, \cite{Baldi20121205}).  This problem of packing tubes and boxes, dealing with heterogeneous items and with distinctive requirements concerning their placement, does not fall into any of the categories tackled in the bibliography; to the best of our knowledge, it is being addressed here for the first time.  Further details on the problem and on the method are conveyed in Section~\ref{sec:problem}.  An illustration of the steps involved in the algorithm is presented in Section~\ref{sec:algandresults}, and conclusions are drawn in Section~\ref{sec:conclusions}.

\section{Problem description and heuristics}
\label{sec:problem}

One of the difficulties of packing in the tube industry is due to the diversity of shapes involved, which calls for different algorithms for each shape.  We will shortly describe each of these algorithms, and then describe the method used for putting all of them together in a single solver.  In the methods we propose, a container (corresponding to the physical object used for packing tubes and boxes inside) is divided into smaller parallelepipeds, which we call \emph{holders}.  The methods are based on long container shapes, which correspond to those that are used in practice.  Tubes are also long, and their placement is always along the depth of the container; please see the left image of Figure~\ref{fig:container3D}.

Orders in this company specify a minimum number of tubes and/or boxes required by the customer.  However, customers are usually interested in using all the available space in the container.  Hence, there is a \emph{value} associated to each tube and box, which is a measure of the interest of inserting additional pieces in the packing.  The first goal is to minimize the number of containers used for packing the specified quantity of tubes and boxes; the second goal is to maximize the monetary value of additional objects (tubes or boxes) which can be packed in the minimum number of containers.

\subsection{Tube packing}
\label{sec:rcpp}

In this industry tubes are usually produced in a continuous extraction machine, and may be cut to different lengths.  A special case is when they are cut to the length of the container inside which they will be shipped.  Before being placed in the container they may be inserted inside other, thicker tubes, so that usage of container space is maximised $-$ a process named \emph{telescoping}.  As in this case all the tubes occupy the full container length, maximising container load is equivalent to maximising the area filled with rings, in a section of the container.  This is called the recursive circle packing problem (RCPP), and has been dealt with in~\cite{pedroso2016itorOCC}.  The idea behind it is to construct packings repeatedly with a greedy method coupled with a random component, and at the end select the best of the solutions encountered.  At each step, tubes may be placed either directly inside the container/holder or telescoped inside other tubes.


For the purpose of this work, two variants of algorithms for the RCPP are used:
\begin{description}
\item[Algorithm T1:] given the dimension of a holder and a set of tubes which do not completely fit there, find a packing with the maximum number of tubes of each diameter, lexicographically, by decreasing diameter.
\item[Algorithm T2:] when the set of tubes completely fits in the holder, determine a packing with minimum height.
\end{description}
In order or decreasing diameter, each tube is tentatively inserted.  In practice, packing larger tubes first is a good heuristic rule, and greedy procedures based on this order usually give satisfactory results.

\subsection{Box packing}
\label{sec:bpp}
Packing of parallelepipeds in a container is a well studied problem, with many different variants.  In our case, boxes are a secondary product; also, in practice it is viable to put boxes on top of tubes but not the inverse.  For these reasons, the relevant box packing variants for our case are:
\begin{description}
\item[Algorithm B1:] given the dimensions of a holder, find the maximum number of boxes from the given input set that can be accommodated inside it, lexicographically, by decreasing volume, or
\item[Algorithm B2:] if all boxes can be placed inside the holder, find the minimum depth required (first goal), and then the minimum height required, keeping the width and depth of the holder fixed.
\end{description}

The method for box placement consists of the following.  A corner is the intersection of three orthogonal planes (sides of previously placed boxes or walls of the container).  The method keeps an updated list of available corners, which is calculated by a process of elimination of points through an analysis of the octants occupied after each placement.  Every corner is assigned a signature that guides the direction of potential placements of a box in that corner.  Each box is placed leaning against the best corner, \ie, the one that leads to minimum depth, height, and width, in lexicographical order.  This is a modified version of the method described in~\cite{Huang2009a,Huang2009b}, including new ideas on using a signature for eliminating octants when updating the corner list.

Placement is always made with the boxes' sides orthogonal to the walls of the container, in contact with container's wall or with previously packed objects in each of the dimensions.  The algorithm attempts placement of each of the boxes in one of six possible orientations, but identical boxes in a row (\ie, fitting next to the previous and in the same orientation) are positioned with the same orientation of the previous, whenever possible.  When the size of a box being packed changes with respect to the previous, each of the six possible orientations is attempted again.  The algorithm is greedy: for each box the best corner is chosen (considering the criteria above), but some diversification is allowed by firstly attempting all the different box sizes that remain to be packed, and then all the possibilities for the orientation of the first box of each type in a row.  The solution selected is the one that leads to the best objective overall.  As the number of different box sizes in an order is typically very small, this method is fairly fast in practice.

\subsection{Method for packing tubes of different lengths}
\label{sec:pack}

The problem when tubes have different lengths becomes much harder to solve than the RCPP, as the solution must consider the 3D disposition of tubes inside the container.  A set of practical rules helps making this problem a little bit better defined:
\begin{itemize}
\item all tubes are placed in parallel to the depth axis;
\item shorter tubes may be placed on top of longer tubes, but not the inverse;
\item more generally, objects placed on top must not protrude with respect to tubes/objects below;
\item boxes may be placed on tubes, but not the inverse.
\end{itemize}

Based on these rules, we propose a heuristic method for dividing a container into holders.  The set of holders inside a container, with sides orthogonal to its walls, form a partition of the container.   Holders' width is fixed to the width of the physical container; this is due to the possibility for the tubes to flow on this dimension.  Furthermore, this allows a simplification in solving the problem, and is very convenient in the industrial setting for communicating the solution.  

Tube holders have a fixed depth, corresponding to the length of the tubes they accommodate.  Tube placement is done by decreasing tube length.  Packing tubes of a given length leads to the division of a holder in three parts, as shown in Figure~\ref{fig:container3D}:
\begin{figure}
  \begin{center}
\begin{tabular}{ m{5cm} m{3mm}  m{5cm} }
    \input{axes.txt}
    & $\rightarrow$ &
    \input{holder1.txt} 
\end{tabular}
  \end{center}
  \caption{A container: depth ($D$), width ($W$) and height ($H$) dimensions, and direction of tube placement (left). Recursive division of a container or holder into two smaller holders, after placement of tubes in the figure at the left.}
\label{fig:container3D}
\end{figure}
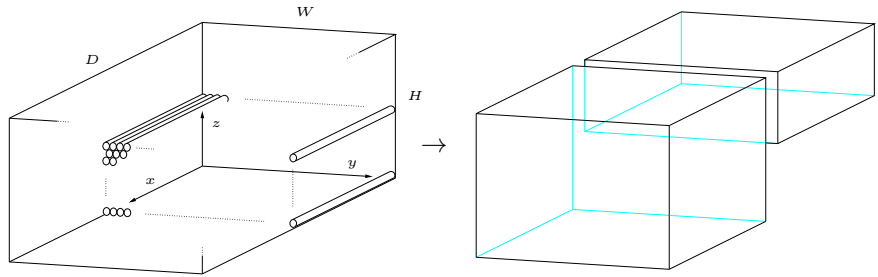
\begin{enumerate}
\item the holder where tubes of the current length are packed, which, having become full, is no longer considered;
\item a holder above it, if some vertical space has been left;
\item a holder beside it, if some horizontal space has been left.
\end{enumerate}
This division is recursive: the holders created this way may be used for an identical packing procedure.  
The height of a holder is determined either by:
\begin{enumerate}
\item the solution of the RCPP with algorithm T2 that minimises the height used for packing the set of tubes assigned to this holder, in case all these tubes fit, or
\item the limit of the physical container:
  \begin{itemize}
  \item when the set of tubes assigned to this holder cannot be packed completely, in which case a new holder with the same length will have to be created afterwards.
  \item when the objects being packed are boxes.
  \end{itemize}
\end{enumerate}

Holders for boxes have a depth identical to the (tube) holder below them; the depth actually used may be smaller, but as boxes are packed after tubes (implying that tube placement could not be done on holders being considered for boxes), unfilled space in box containers is not usable.

Whenever there is a non-empty list of items to pack and the list of holders where items could possibly fit is empty, a new physical container is added, and a holder with its size is created.  These ideas are structured in Algorithm~\ref{alg:main}.  
\begin{algorithm}[h!tb]
  \begin{footnotesize}
    \DontPrintSemicolon
    \SetKwFunction{algo}{algo}\SetKwFunction{fill}{fill}
    \KwData{instance:
      \begin{itemize}
      \item lists of tubes $T$ and boxes $B$ to pack;
      \item physical container's width $W$, height $H$,  and depth $D$;
      \end{itemize}
    }
    \KwResult{
      \begin{itemize}
      \item list of containers required for packing;
      \item list of holders in each container and their position inside the container;
      \item list of tubes or boxes in each holder and their position inside the holder.
      \end{itemize}
    }
    $C \eq $ list of containers in the solution (initially empty)\;  \label{line:begin-init}
    \While{$T$ or $B$ are not empty}{
      $c \eq $ empty holder with dimensions $W \times H \times D$ \;  \label{line:init-container}
      $c \eq $ \fill{$D, W, H, c, T, B$}\;  \label{line:init-fill}
      append $c$ to $C$\;  \label{line:end-init}
    }
    \Return{$C$}\;
    ~\\
    \SetKwProg{myproc}{Procedure}{}{}
    \myproc{\fill{$D, W, H, c, T, B$}}{
      $d \eq 0$ \tcp*{for keeping track of the amount of holder depth used}
      \Repeat{$U$ is empty}{ \label{line:repeat}
        $U \eq $ tubes from $T$ with length $\leq D - d$\;
        \If(\tcp*[f]{no tubes fit; fill with boxes and return}){$U$ is empty}{  \label{line:start-boxes}
          $X \eq $ boxes from $B$ with one or more sides length $\leq D - d$\;
          \If{$X$ is empty}{
            \Return{};
          }
          $h \eq $ holder with dimensions $W \times H \times (D - d)$ filled with boxes from $X$\;
          insert $h$ in $c$\;
          $B \eq$ boxes remaining unpacked\;  \label{line:endboxes}
          \Return{};
        }
        $\ell \eq $ maximum tube length not larger than $D - d$ \tcp*{create a slice of depth $\ell$}  \label{line:slice}
        $U_\ell \eq $ list of tubes of $U$ with length $\ell$\;
        $h \eq $ holder with dimensions $W \times H \times \ell$ filled with tubes $U_\ell$\;  \label{line:new-tube-holder}
        \If{all tubes $U_\ell$ were inserted}{
          update height of holder $h$\;
        }
        $a \eq $ height of holder $h$\;
        insert $h$ in $c$\;
        $T \eq$ tubes remaining unpacked\;
        \If{some tubes or boxes fit in $W \times (H-a) \times \ell$}{
          \fill{$\ell, W, H-a, c, T, B$}\; \label{line:rec}
        }
        $d \eq d + \ell$\; \label{line:update-d}
      }
      \Return{};
    }
  \end{footnotesize}
    \caption{Recursive algorithm for tube and box packing.}
    \label{alg:main}
\end{algorithm}
The input for this algorithm is the list of tubes and boxes to pack, each characterised by its dimensions, and the size of the containers where these items are to be packed.  The output is a list of containers required for packing all the items, divided into holders.  Each holder is characterised by its dimensions and position in the container, and includes a list of items (either tubes or boxes) packed inside it.  Each of these items, in turn, is characterised by its position inside the holder.  The main algorithm calls a recursive function \verb|fill|, which receives as input a holder (initially empty, with the dimensions of the physical container) and fills it with tubes and boxes available.  This function works by creating a transversal slice of the holder with the length of the longest tubes that still fit in it (lines \ref{line:slice} to~\ref{line:update-d}), originating a tube holder.  If tubes of that length completely fill this slice, the algorithm moves to the next slice.  Otherwise, a holder is created above these tubes, using the remaining space in the slice, and yet unpacked tubes and boxes are inserted there (line~\ref{line:rec}); this is the recursive part of the algorithm.  When no more tubes fit in a slice, a holder for boxes is created and filled (lines \ref{line:start-boxes} to~18). 

\section{Algorithm's execution and results}
 \label{sec:algandresults}

In this section we will describe each of the steps involved in using the algorithm.  For this illustration, we have selected an instance found in practice; the actual data is listed in~\ref{sec:appendix}.

\subsection{Data}
\label{sec:data}

An instance of this problem defines container's dimensions, as well as the tubes and boxes to be packed.  All containers are assumed to have identical dimensions; in practice this may not be true, as some discrepancies exist, but this information is usually unknown in advance.  However, the methods proposed do work straightforwardly if different dimensions to be used are given as an ordered list.

Tubes are characterised by external and internal diameters and by their length.  Typically, orders include many identical tubes; hence, the number of tubes of a given dimension is also a datum.  Real tubes may have a socket in one side; in this case, to ensure that the selected tube will actually fit in the holder, the external diameter of the tube is updated to the socket's external diameter.  Tubes to be packed are kept in a list ordered by decreasing length.  Tubes of the same length are ordered by decreasing external diameter.

Boxes are characterised by width, height, and depth; as for tubes, the number of boxes of each type is also a datum.  In many situations, rotation of boxes is only allowed on the vertical axis; in this industry, however, boxes are typically very light and can be safely rotated in any direction.  Boxes to be packed are kept in a list ordered by decreasing volume.  


\subsection{Algorithm's execution: a commented example}
\label{sec:execution}

\begin{figure}[h!tbp]
  \centering
  \includegraphics[width=.95\textwidth]{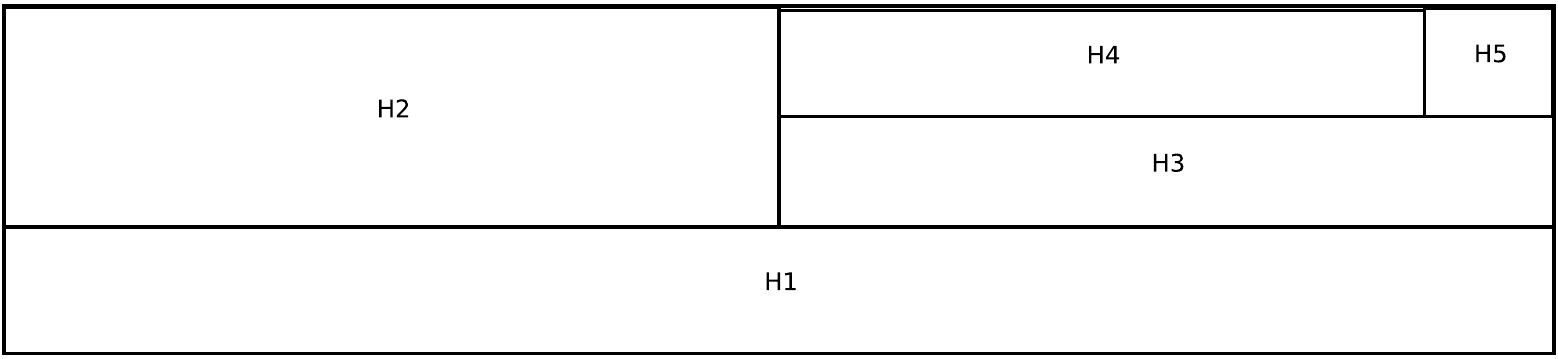}~\\~\\
  \includegraphics[width=.95\textwidth]{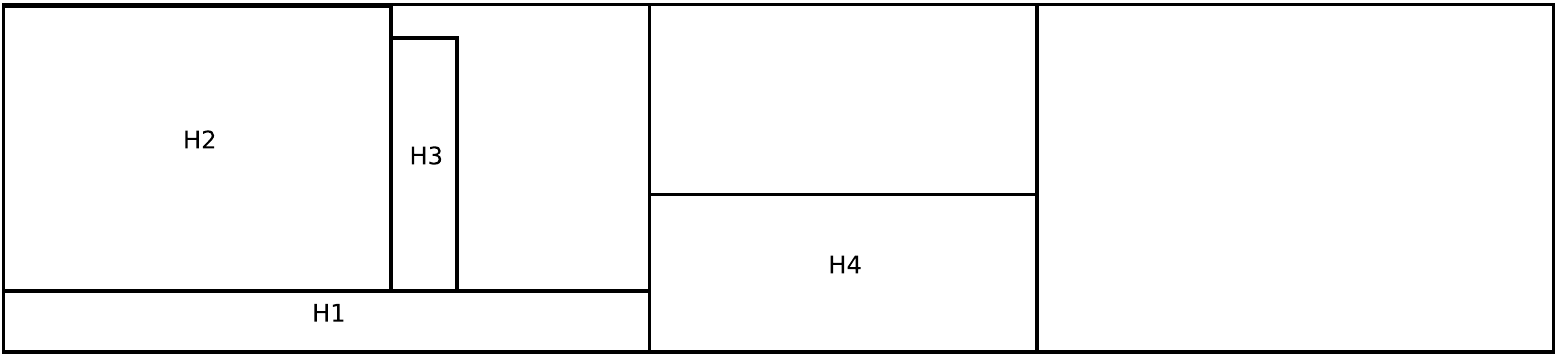}
\caption{Longitudinal sections of the load of Container 1 (top) and Container 2 (bottom) in the solution obtained for the instance presented in~\ref{sec:appendix}.}
\label{fig:solH}       
\end{figure}

\begin{figure}[t]
  \centering
  \includegraphics[width=.49\textwidth]{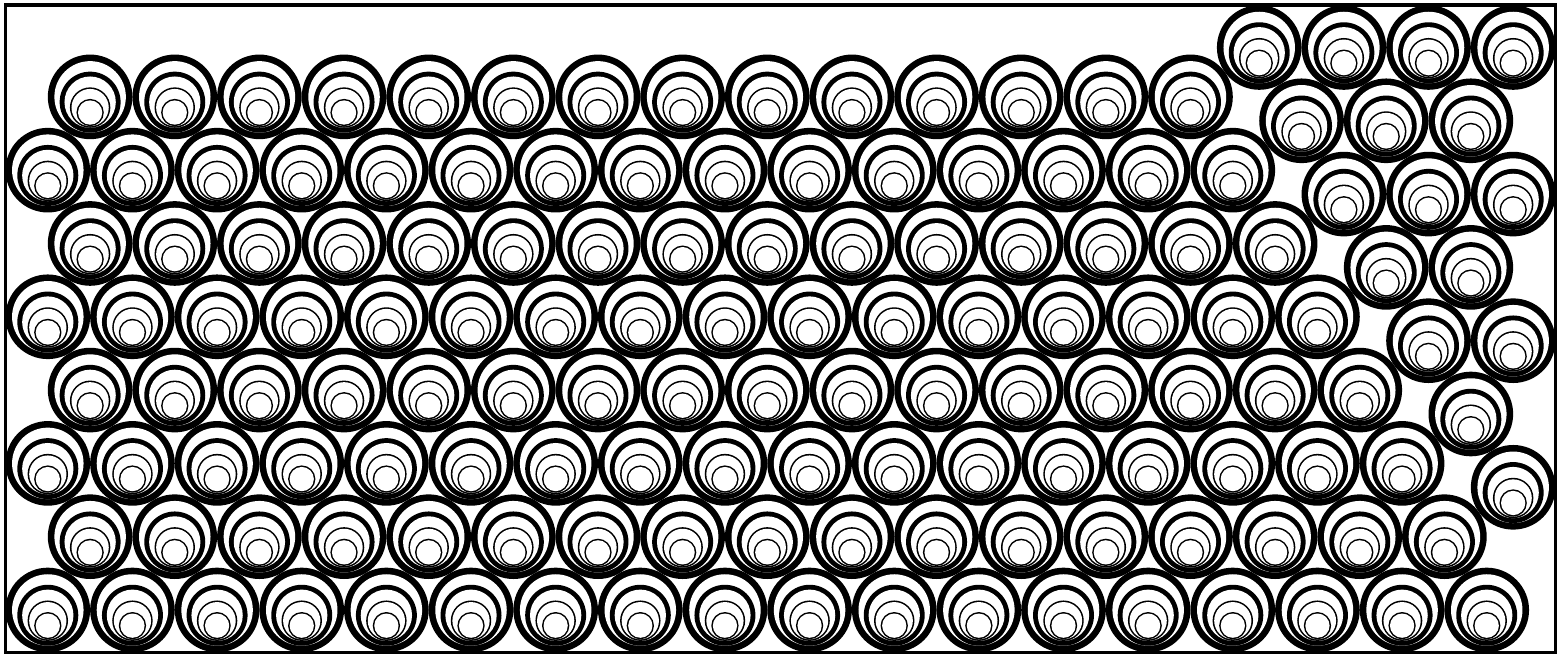}
  \includegraphics[width=.49\textwidth]{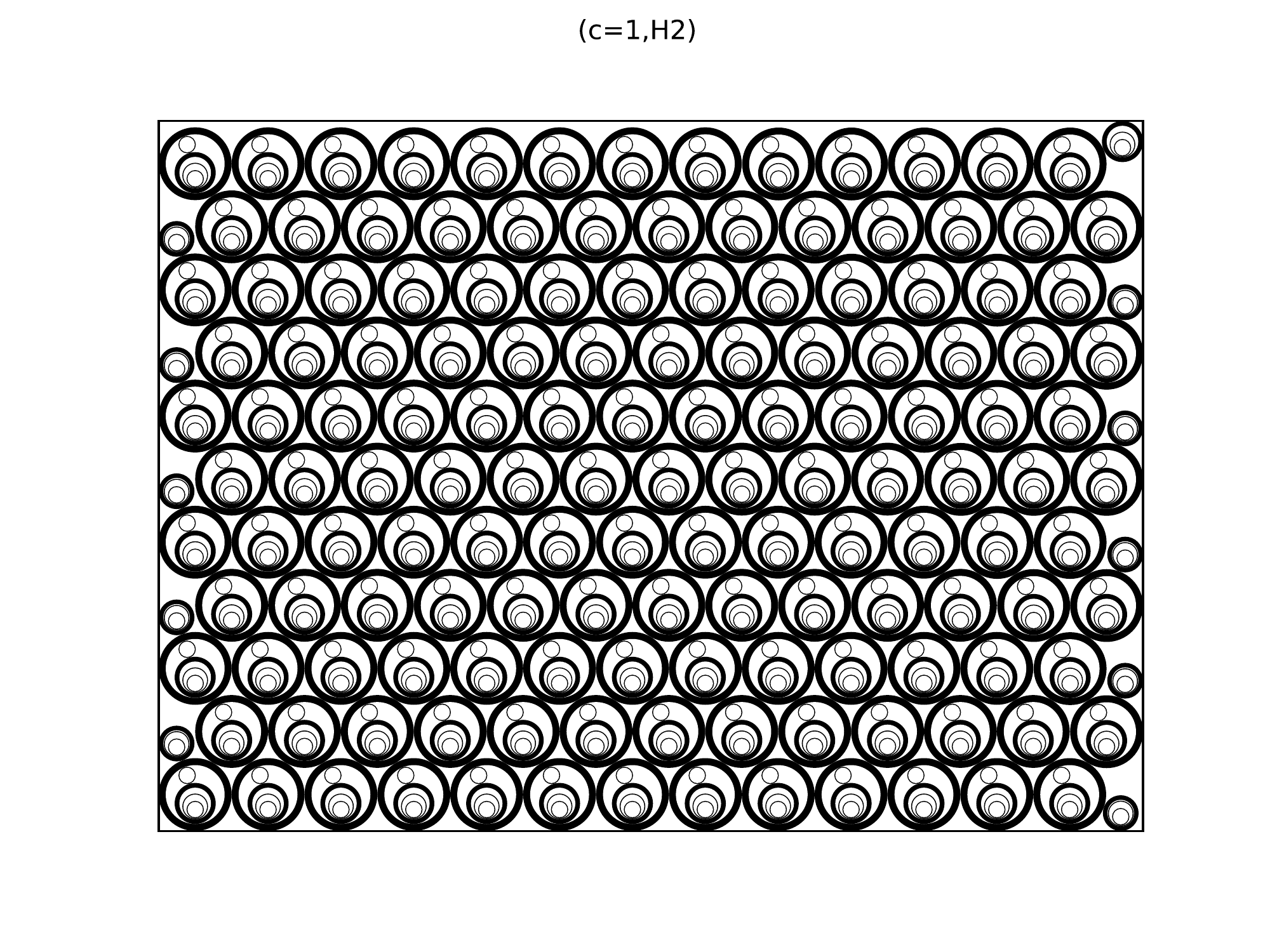}\\
  \includegraphics[width=.49\textwidth]{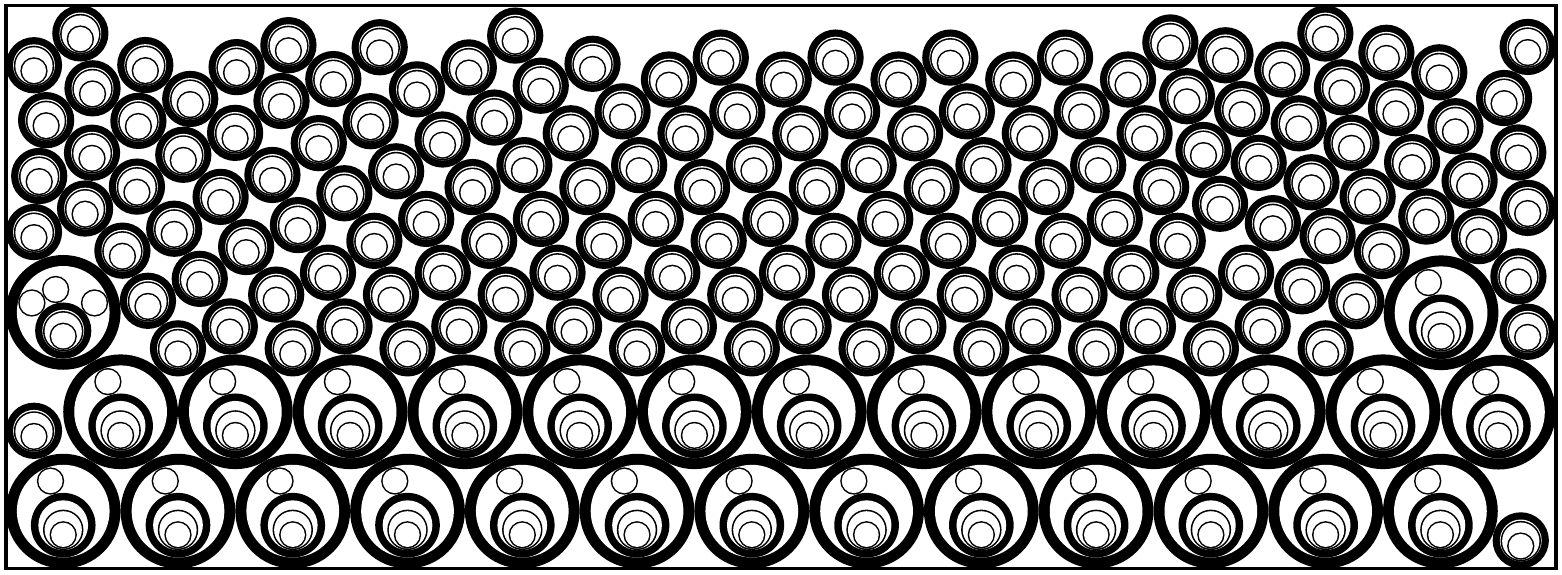}
  \includegraphics[width=.49\textwidth]{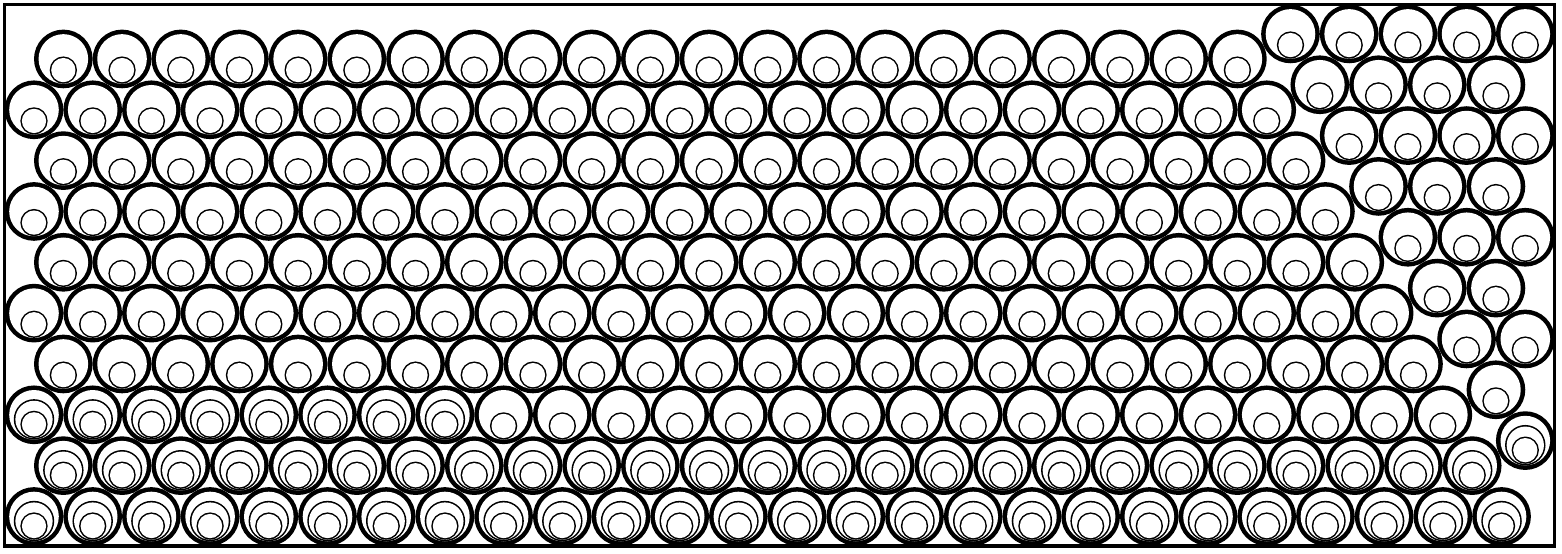}
\caption{Transversal sections of tube holders for Container 1: H1 (top-left) to H4 (bottom-right).}
\label{fig:solH1}       
\end{figure}

The first steps of execution of Algorithm~\ref{alg:main} consist of adding to the solution a new empty container, whose dimensions are used as a holder's size.  Tubes and boxes will be packed in this holder by means of procedure \verb|fill| (line~\ref{line:init-fill}); in the initial call, the longest tubes have the same length as the container, and are assigned to holder \verb|H1|, placed at the bottom of Container~1, as shown in Figure~\ref{fig:solH} (top).  The disposition of tubes in holder \verb|H1| is determined in line~\ref{line:new-tube-holder}, by calling Algorithm T2; it is shown in Figure~\ref{fig:solH1} (top-left)\footnote{Notice that after having filled tubes of the current length, shorter tubes are tentatively telescoped inside those previously packed; for clarity, this step is omitted in Algorithm~\ref{alg:main}.}.  When the algorithm reaches line~\ref{line:rec}, the space above \verb|H1| is used for creating new holders, through a recursive call to procedure \verb|fill|.  In this call, in the first iteration of the cycle in line~\ref{line:repeat}, tubes of the current length completely fill the vertical space in \verb|H2|; hence, the maximum number of tubes is placed there by Algorithm T1, and there is no recursive call in line~\ref{line:rec}.  In the second iteration, holder \verb|H3| is created for tubes of the same length (line~\ref{line:new-tube-holder}).  This time space is left above it, which is used in a new recursive call, originating holders \verb|H4| and \verb|H5|; the latter, having no tubes that fit in, is used for packing boxes (lines~\ref{line:start-boxes} to~18), using Algorithm B1. 
At this point, no usable space is left in the holders created so far; hence, the recursion is complete and returns to line~\ref{line:init-fill}, and Container~1 is closed.

As there are still tubes and boxes awaiting to be packed, a new container is added in line~\ref{line:init-container}.  The recursive call to procedure \verb|fill| is similar to the previous, except that now all the remaining boxes can be packed in holder~\verb|H3| (using Algorithm B2), and hence some space beside it is left unused.  The same for the space on top of and beside \verb|H4|.  Most likely, being shown this solution, the decision maker would inform the customer that space is available in a container, ask if some additional product should be dispatched, and call Algorithm~\ref{alg:main} again.

It should be noted that the problems being tackled by algorithms T1, T2, B1, B2 are very difficult to solve optimally, and only an approximate solution can be found in a reasonable time.  The total time intended to be used for finding a solution is input when starting Algorithm~\ref{alg:main}, and is split among all the algorithms it calls by a heuristic rule.

\begin{figure}[t]
  \centering
  \begin{minipage}[top]{.49\textwidth}
  \includegraphics[width=1.\textwidth]{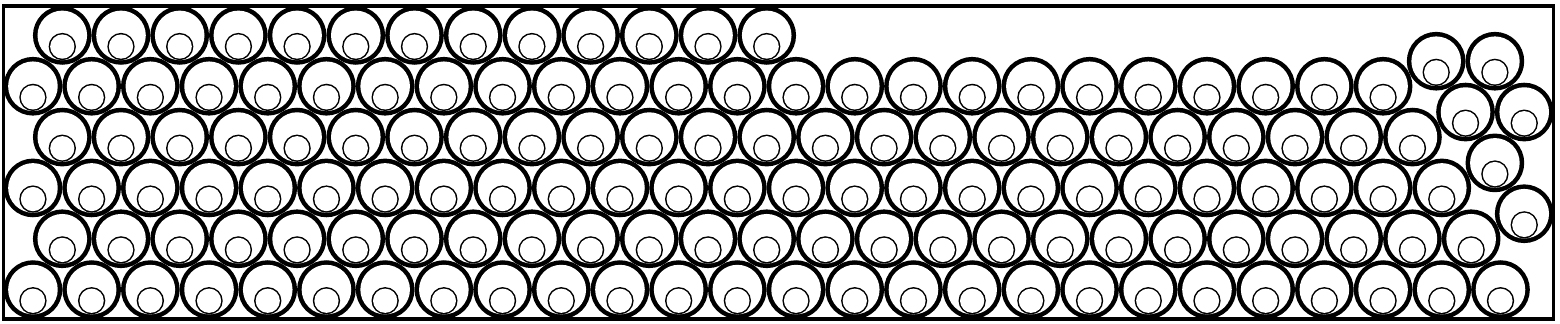}\\
  \includegraphics[width=1.\textwidth]{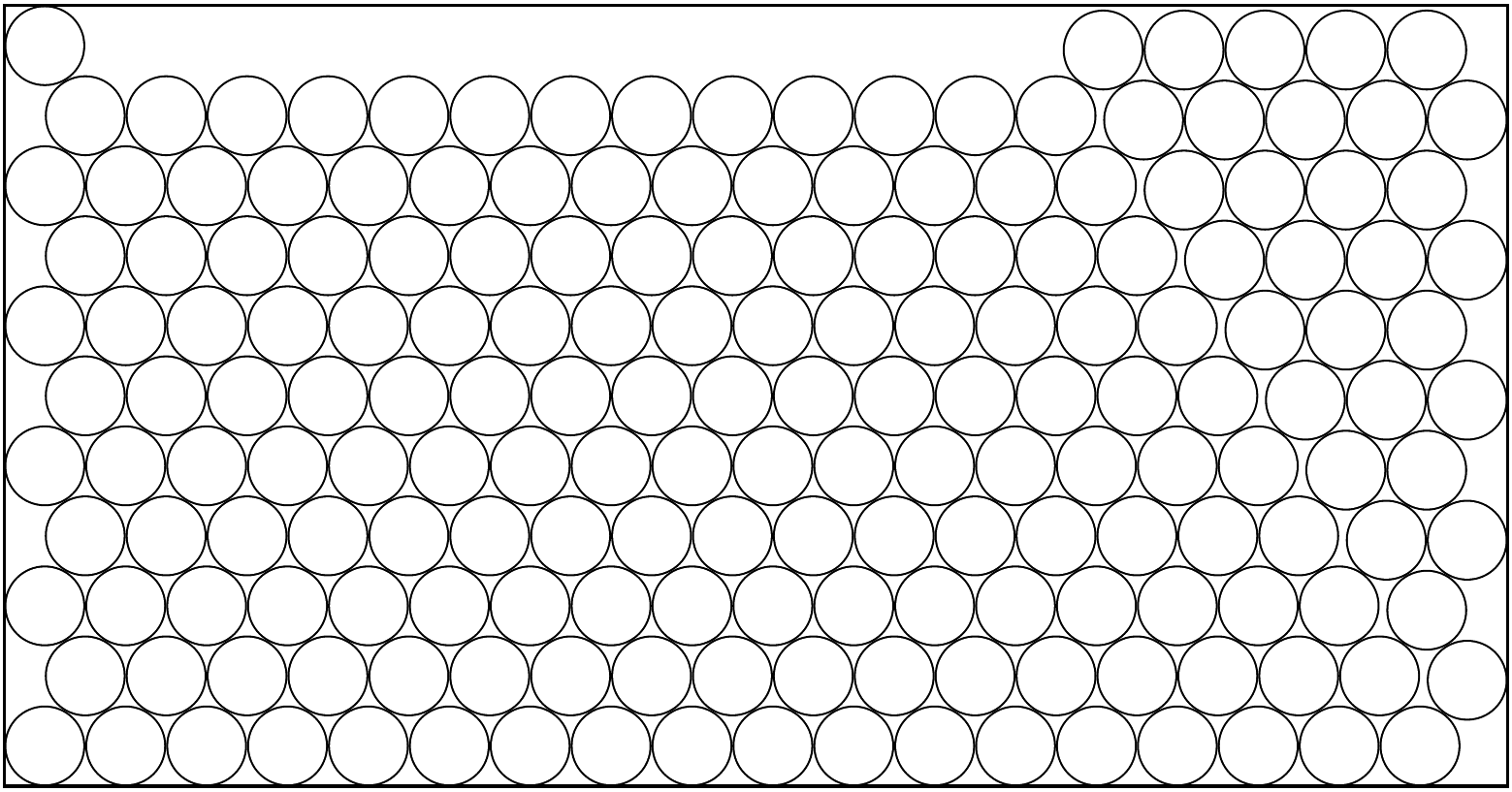}
  \end{minipage}
  \begin{minipage}[top]{.49\textwidth}
  \includegraphics[width=1.\textwidth]{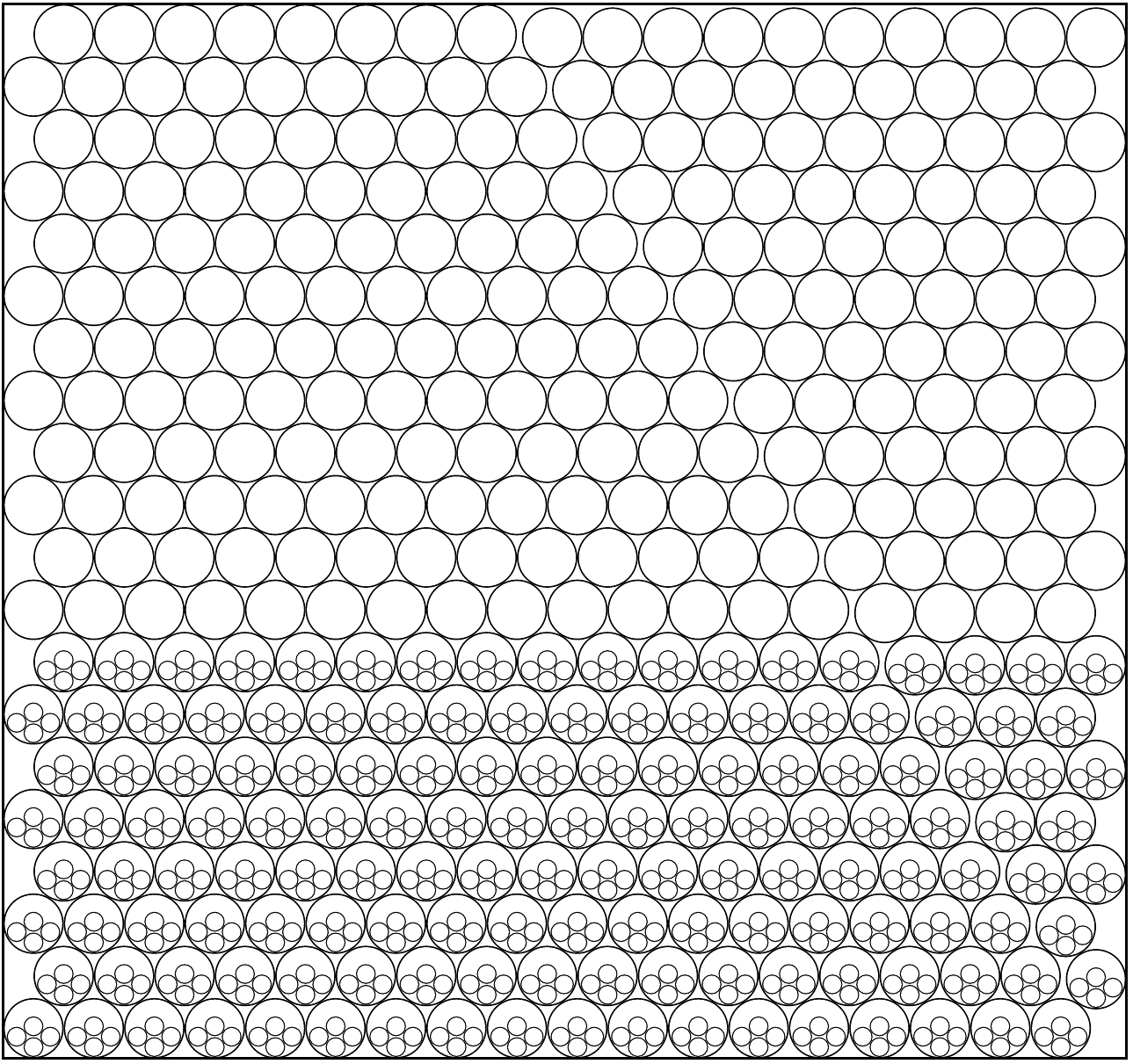}
  \end{minipage}
\caption{Transversal sections of tube holders for Container 2: H1 (top-left), H4, and H2 (right).}
\label{fig:solH2}       
\end{figure}

\subsection{Solution communication}
\label{sec:communic}

The basis for communicating the solution at an operational level are representations of the container's division into holders, as in Figure~\ref{fig:solH}.  This layout is followed closely at the time of actual packing.  However, the disposition of tubes shown in figures \ref{fig:solH1} and \ref{fig:solH2} is only followed for very wide tubes, which may be tricky to place in the container.  Narrow tubes (as those shown) usually find a near-optimal configuration when placed arbitrarily.  Hence, algorithms T1 and T2 are used mostly for determining quantities of tubes that can be dispatched.  Diagrams representing the disposition of boxes are also awkward to visualise; instead, the software prepares a list of boxes packed in each holder, together with their orientation.

\subsection{Results}
\label{sec:results}

To our knowledge, this problem has not been tackled previously in the literature.  Even though there are no methods for comparing the performance of our algorithm, we have prepared a set of benchmark instances that can be used for that purpose; the set is available as data associated to this paper, in~\cite{pedroso2016occdata}.

This set of instances helps assessing the CPU time used, as well as the size of relevant instances for the company. For filling each holder, we are using simple construction heuristics for packing tubes or boxes.  Each of the instances completely fills the container; an additional unit of any the items involved would imply using an additional container (by using the simplest heuristics for algorithms T1, T2, B1, B2; these solutions may potentially be improved by allowing more CPU time, or by using more sophisticated methods for box and tube packing in each of the holders).  Hence, the value of additional items packed in each of the solutions (\ie, the value of second goal) is zero.
\begin{table}[htbp]
  \centering
  \begin{tabular}{l|rrrr|rr}
    Instance &  N &    T & M &   B & C & CPU \\ \hline
    occ-cie-1    &  3 &  832 & 2 &  21 & 1 &   3 \\
    occ-cie-2    &  3 & 2502 & 2 & 114 & 3 &  18 \\
    occ-cie-3    &  5 & 1542 & 2 &  54 & 1 &  10 \\
    occ-cie-4    &  5 & 2798 & 2 & 150 & 2 &  26 \\
    occ-cie-5    &  9 & 3014 & 3 &  53 & 1 &   9 \\
    occ-cie-6    &  9 & 4326 & 3 &  17 & 2 &   5\\
    occ-cie-7    & 10 &  359 & 3 &  27 & 1 &   5\\
    occ-cie-8    & 10 & 1551 & 3 & 125 & 4 &  21\\
  \end{tabular}
  \caption{Characteristics of instances and CPU time required for a basic heuristic solution.  N is the number of different tubes, T is the total number of tubes, M is the number of different boxes, B is the total number of boxes, C is the number of containers used, and CPU is the time (in seconds) required for obtaining the solution.}
\end{table}


These results show that the runtime required is acceptable.  By visually inspecting the solutions obtained, we could observe that the usage of the container was very good, even though a potential for improvement is perceptible.

\label{tab:results}
\section{Discussion and conclusions}
\label{sec:conclusions}

This paper presents a heuristic method for packing tubes and boxes together in containers, a problem arising in the tube industry.  This problem is very hard, as its components involve the solution of other known difficult problems.  Arising in a key area for pricing and delivery logistics, its solution is relevant in practice.  Not all practical problems are suitable to a clean, standard formulation in mathematical programming.  Here, even the packing subproblems (recursive circle packing, box packing) have been proven NP-hard, and are known to be extremely difficult to solve in practice.  Hence, we had to resort the an approximative method, designed specifically to this problem.

Some rules for packing previously established in the company helped to keep focus on important parts of the method, and at the same time somehow simplify it so as to be able to achieve realistic solutions.  The method consists of a procedure for a three-dimensional partition of the container into holders, and the main design decision has been to impose that holders --- divisions of the container --- all have the same width as the physical container.  This division is recursive, and is driven by tube packing: tubes are positioned in parallel to the container's depth axis, by decreasing length, originating holders above and beside them.  Holders can be plotted as rectangles in a diagram corresponding to a vertical, longitudinal section of the container.  Each holder is characterised by its position and by the set of tubes or boxes that it will carry.

As the products involved in this industry are relatively inexpensive, have a low weight to volume ratio, and often have to travel long distances, logistic costs in delivery are considerable; their optimisation is a major factor for competitiveness.  The proposed method provides good, practical solutions in a very limited time, as required for its real-world application.  

To the best of our knowledge, there are no other methods dealing with this problem; therefore, we cannot compare the performance of our method to others.  We can, however, provide an empirical assessment made by the company, putting in evidence the savings as compared to previous solutions.  These are the improvements in terms of reduction of the number of containers that would have been used to send historical orders, and in terms of the increase in weight per container that could be have been sent:
\begin{center}
\begin{tabular}{lrr}
        & Containers & Weight   \\ \hline
Minimum & 6.3\%      & 8.3\%    \\ 
Maximum & 15.8\%     & 22.5\%   \\ 
Average & 11.2\%     & 15.4\%   \\ 
\end{tabular}
\end{center}
However, the main benefits seen by the company are on the flexibility on management that the method allows.

Dealing with boxes and tubes simultaneously allowed filling spaces that, previously, were unexploited.  Solutions obtained ensure that at the time of loading there is a feasible sequence for the operations involved: filling bottom up, from the rear to the door of the container.  Our procedure allowed improving quality of service and communication with customers, and its usage was considered an important factor for reductions in logistic costs and for facilitating timely responses to order enquiries.  

There are several interesting directions for future work.  The greedy approach used for choosing tube length at each step may be replaced by better alternatives, for example using metaheuristics or tree search.  A more practical direction consists of taking into account both load weight and the resistance of tubes and boxes for determining what can be stacked on top of what.  
An approach based on mathematical programming formulations is unlikely to provide practical results, as the subproblems involved are known to be difficult even when tackled separately, but it would undoubtedly be theoretically interesting.  
Given the difficulty and the rather specific nature of this problem, we believe that currently there is no need of benchmark instances, as it is unlikely that other approaches for tackling it emerge; but this may change in the future.
Finally, considering the possibility of placing tubes/cylinders in any orthogonal direction is also interesting and relevant in practice.

\paragraph{Acknowledgements}
We would like to thank S{\'\i}lvia Cunha for her help in coding part of the algorithms.  
This work was partially supported by FEDER, QREN e Compete, within project \emph{OCC: Optimizador de Carga em Contentores, QREN SII\&DT Projeto em Co-Promo{\c c}{\~a}o  13824}, and includes contributions from its research team.  This work is partly funded by FCT – Funda{\c c}{\~a}o para a Ci{\^e}ncia e a Tecnologia (Portuguese Foundation for Science and Technology) within project UID/EEA/50014/2013.


\appendix
\section{Data used in the illustration}
\label{sec:appendix}

\begin{small}
\begin{verbatim}
container
# len   height  depth
2350    2690    12000

tubes
# tubes data, where:
#   ID - code
#   idiam - internal diameter
#   ediam - external diameter
#   len - tube length
#   number - number of tubes in the order
# ID    idiam   ediam   len     number
A       108.38  128.33  12000   144
B       77.8    92.42   12000   144
C       140.59  174     6000    171
D       75.39   98      6000    171
E       63.39   85.30   6000    171
F       120.20  126.46  3000    546
G       56.50   61.31   3000    546
H       36.80   41.41   3000    1638
I       75.2    89      5000    403

boxes
# ID    width   height  depth   number
B1      500     330     800     10       
B2      460     200     700     10
B3      420     310     600     10     
\end{verbatim}
  
\end{small}

\end{document}

%% file: axes.txt
\begin{picture}(0,0)%
\includegraphics{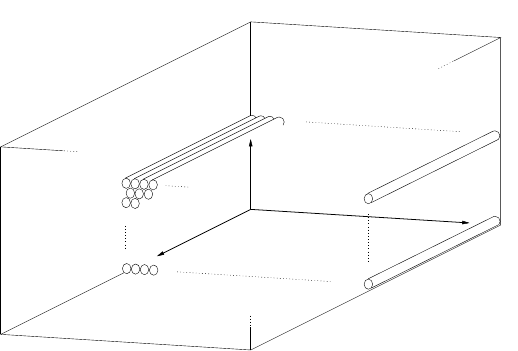}%
\end{picture}%
\setlength{\unitlength}{987sp}%
\begingroup\makeatletter\ifx\SetFigFont\undefined%
\gdef\SetFigFont#1#2#3#4#5{%
  \reset@font\fontsize{#1}{#2pt}%
  \fontfamily{#3}\fontseries{#4}\fontshape{#5}%
  \selectfont}%
\fi\endgroup%
\begin{picture}(9927,6705)(1189,-7873)
\put(11101,-3511){\makebox(0,0)[lb]{\smash{{\SetFigFont{5}{6.0}{\familydefault}{\mddefault}{\updefault}{\color[rgb]{0,0,0}$H$}%
}}}}
\put(4576,-5611){\makebox(0,0)[lb]{\smash{{\SetFigFont{5}{6.0}{\familydefault}{\mddefault}{\updefault}{\color[rgb]{0,0,0}$x$}%
}}}}
\put(6226,-4186){\makebox(0,0)[lb]{\smash{{\SetFigFont{5}{6.0}{\familydefault}{\mddefault}{\updefault}{\color[rgb]{0,0,0}$z$}%
}}}}
\put(9601,-5161){\makebox(0,0)[lb]{\smash{{\SetFigFont{5}{6.0}{\familydefault}{\mddefault}{\updefault}{\color[rgb]{0,0,0}$y$}%
}}}}
\put(3076,-2611){\makebox(0,0)[lb]{\smash{{\SetFigFont{5}{6.0}{\familydefault}{\mddefault}{\updefault}{\color[rgb]{0,0,0}$D$}%
}}}}
\put(8326,-1411){\makebox(0,0)[lb]{\smash{{\SetFigFont{5}{6.0}{\familydefault}{\mddefault}{\updefault}{\color[rgb]{0,0,0}$W$}%
}}}}
\end{picture}%

%% file: holder1.txt
\begin{picture}(0,0)%
\includegraphics{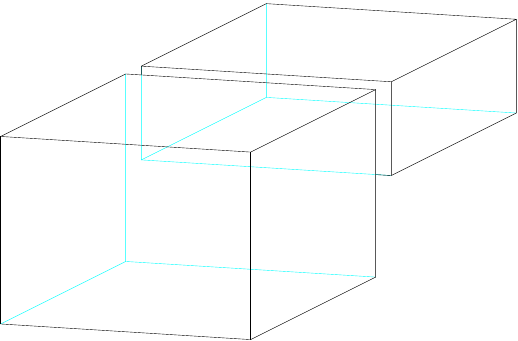}%
\end{picture}%
\setlength{\unitlength}{987sp}%
\begingroup\makeatletter\ifx\SetFigFont\undefined%
\gdef\SetFigFont#1#2#3#4#5{%
  \reset@font\fontsize{#1}{#2pt}%
  \fontfamily{#3}\fontseries{#4}\fontshape{#5}%
  \selectfont}%
\fi\endgroup%
\begin{picture}(9924,6474)(1189,-7873)
\end{picture}%

%% file: DCC_occ-paper.bbl
\begin{thebibliography}{1}
\providecommand{\url}[1]{{#1}}
\providecommand{\urlprefix}{URL }
\expandafter\ifx\csname urlstyle\endcsname\relax
  \providecommand{\doi}[1]{DOI~\discretionary{}{}{}#1}\else
  \providecommand{\doi}{DOI~\discretionary{}{}{}\begingroup
  \urlstyle{rm}\Url}\fi

\bibitem{Baldi20121205}
Baldi, M.M., Crainic, T.G., Perboli, G., Tadei, R.: The generalized bin packing
  problem.
\newblock Transportation Research Part E: Logistics and Transportation Review
  \textbf{48}(6), 1205 -- 1220 (2012)

\bibitem{Crainic2009744}
Crainic, T.G., Perboli, G., Tadei, R.: Ts2pack: A two-level tabu search for the
  three-dimensional bin packing problem.
\newblock European Journal of Operational Research \textbf{195}(3), 744 -- 760
  (2009)

\bibitem{Huang2009b}
Huang, W., He, K.: A caving degree approach for the single container loading
  problem.
\newblock European Journal of Operational Research \textbf{196}(1), 93--101
  (2009)

\bibitem{Huang2009a}
Huang, W., He, K.: A new heuristic algorithm for cuboids packing with no
  orientation constraints.
\newblock Comput. Oper. Res. \textbf{36}(2), 425--432 (2009)

\bibitem{Lim2003471}
Lim, A., Rodrigues, B., Wang, Y.: A multi-faced buildup algorithm for
  three-dimensional packing problems.
\newblock Omega \textbf{31}(6), 471 -- 481 (2003)

\bibitem{Lodi2002410}
Lodi, A., Martello, S., Vigo, D.: Heuristic algorithms for the
  three-dimensional bin packing problem.
\newblock European Journal of Operational Research \textbf{141}(2), 410 -- 420
  (2002)

\bibitem{pedroso2016occdata}
Pedroso, J.P.: A dataset for the problem of packing tubes and boxes in a
  container.
\newblock Internet repository, version 1.0 (2016).
\newblock \url{http://www.dcc.fc.up.pt/~jpp/code/occ}

\bibitem{pedroso2016itorOCC}
Pedroso, J.P., Cunha, S., Tavares, J.N.: Recursive circle packing problems.
\newblock International Transactions in Operational Research \textbf{23}(1-2),
  355--368 (2016)

\end{thebibliography}
